\documentclass[english, twocolumn, 10pt, aps, superscriptaddress, floatfix, prb, citeautoscript]{revtex4-2}
\pdfoutput=1
\usepackage[utf8]{inputenc}
\usepackage[T1]{fontenc}
\usepackage{verbatim}
\usepackage{mathtools}
\usepackage{amsmath}
\usepackage{amssymb}
\usepackage{graphicx}
\usepackage{wasysym}
\usepackage{layouts}
\usepackage{siunitx}
\usepackage{bm}
\usepackage{xcolor}
\usepackage[colorlinks, citecolor={blue!50!black}, urlcolor={blue!50!black}, linkcolor={red!50!black}]{hyperref}
\usepackage{bookmark}
\usepackage{tabularx}
\usepackage{microtype}
\usepackage{babel}
\hypersetup{pdfauthor={Madeleine Geers, David M. Jarvis, Cheng
Liu, Siddharth S. Saxena, Jem Pitcairn, Emily Myatt, Sebastian A.
Hallweger, Silva M. Kronawitter, Gregor Kieslich, Sanliang Ling, Andrew
B. Cairns, Dominik Daisenberger, Oscar Fabelo, Laura
Cañadillas-Delgado, Matthew J. Cliffe},pdftitle={High pressure behaviour
of the magnetic van der Waals molecular framework Ni(NCS)\(_2\)}}
\usepackage[version=3]{mhchem}

\DeclarePairedDelimiter\abs{\lvert}{\rvert}
\DeclarePairedDelimiter\norm{\lVert}{\rVert}

\makeatletter
\let\oldabs\abs
\def\abs{\@ifstar{\oldabs}{\oldabs*}}
\let\oldnorm\norm
\def\norm{\@ifstar{\oldnorm}{\oldnorm*}}
\makeatother

\newcolumntype{L}[1]{>{\raggedright\arraybackslash}p{#1}}
\newcolumntype{C}[1]{>{\centering\arraybackslash}p{#1}}
\newcolumntype{R}[1]{>{\raggedleft\arraybackslash}p{#1}}

\renewcommand{\citep}{\cite}

\begin{document}

\title{High pressure behaviour of the magnetic van der Waals molecular
framework Ni(NCS)\(_2\)}

\author{Madeleine Geers}
\affiliation{School of Chemistry, University Park, University of
Nottingham, Nottingham, NG7 2RD, United Kingdom}
\affiliation{Institut Laue Langevin, 71 avenue des Martyrs CS 20156,
38042 Grenoble Cedex 9, France}
\author{David M. Jarvis}
\affiliation{Cavendish Laboratory, University of Cambridge, Cambridge,
CB3 0HE, United Kingdom}
\author{Cheng Liu}
\affiliation{Cavendish Laboratory, University of Cambridge, Cambridge,
CB3 0HE, United Kingdom}
\author{Siddharth S. Saxena}
\affiliation{Cavendish Laboratory, University of Cambridge, Cambridge,
CB3 0HE, United Kingdom}
\author{Jem Pitcairn}
\affiliation{School of Chemistry, University Park, University of
Nottingham, Nottingham, NG7 2RD, United Kingdom}
\author{Emily Myatt}
\affiliation{School of Chemistry, University Park, University of
Nottingham, Nottingham, NG7 2RD, United Kingdom}
\author{Sebastian A. Hallweger}
\affiliation{TUM Natural School of Sciences, Technical University of
Munich, D-85748 Garching, Germany}
\author{Silva M. Kronawitter}
\affiliation{TUM Natural School of Sciences, Technical University of
Munich, D-85748 Garching, Germany}
\author{Gregor Kieslich}
\affiliation{TUM Natural School of Sciences, Technical University of
Munich, D-85748 Garching, Germany}
\author{Sanliang Ling}
\affiliation{Advanced Materials Research Group, Faculty of Engineering,
University of Nottingham, University Park, Nottingham NG7 2RD, United
Kingdom}
\author{Andrew B. Cairns}
\affiliation{Department of Materials, Imperial College London, Royal
School of Mines, Exhibition Road, SW7 2AZ, United Kingdom}
\author{Dominik Daisenberger}
\affiliation{Diamond Light Source, Chilton, Didcot OX11 0DE, United
Kingdom}
\author{Oscar Fabelo}
\affiliation{Institut Laue Langevin, 71 avenue des Martyrs CS 20156,
38042 Grenoble Cedex 9, France}
\author{Laura Cañadillas-Delgado}
\email[Electronic address: ]{canadillas-delgado@ill.fr}
\affiliation{Institut Laue Langevin, 71 avenue des Martyrs CS 20156,
38042 Grenoble Cedex 9, France}
\author{Matthew J. Cliffe}
\email[Electronic address: ]{matthew.cliffe@nottingham.ac.uk}
\affiliation{School of Chemistry, University Park, University of
Nottingham, Nottingham, NG7 2RD, United Kingdom}

\date{\today}

\begin{abstract}
\noindent Two-dimensional materials offer a unique range of magnetic,
electronic and mechanical properties which can be controlled by external
stimuli. Pressure is a particularly important stimulus, as it can be
achieved readily and can produce large responses, especially in
low-dimensional materials. In this paper we explore the
pressure-dependence of the structural and magnetic properties of a
two-dimensional van der Waals (vdW) molecular framework antiferromagnet
with ferromagnetic layers, \ce{Ni(NCS)2}, up to \(8.4\) kbar. Through a
combination of X-ray and neutron diffraction analysis, we find that
\ce{Ni(NCS)2} is significantly more compressible than comparable vdW
metal halides, and its response is anisotropic not only out of the
plane, but also within the layers. Using bulk magnetisation and neutron
diffraction data, we show that the ambient layered antiferromagnetic
phase is maintained up to the largest investigated pressure, but with an
enhanced Néel temperature, \(T_\mathrm{N}\),
(\(\Delta T_\mathrm{N} / T_\mathrm{N} = +19\) \%) and a large pressure
sensitivity
(\(Q = \frac{1}{T_\mathrm{N}} \frac{\mathrm{d}T_\mathrm{N}}{\mathrm{d}P} = +2.3\)
\% kbar\(^{-1}\)), one of the larger values of magnetic pressure
responsiveness for a vdW material. Density functional theory
calculations suggest that this is due to increasing
three-dimensionality. These results provide some of the first insights
into the pressure response of molecular framework vdW magnets and
suggest investigation of other molecular framework vdW magnets might
uncover contenders for future pressure-switchable devices.
\end{abstract}

\flushbottom
\maketitle

\hypertarget{introduction}{%
\section{Introduction}\label{introduction}}

Magnetic van der Waals (vdW) materials, compounds formed from
two-dimensional layers held together through weak dispersion
interactions, have been the subject of much recent interest as potential
single-layer magnetic materials capable of being embedded in spintronic
devices.\citep{cortie_twodimensional_2020} The weak interactions between
layers means pressure can switch the sign of interactions, for example
bilayer \ce{CrI3} undergoes an antiferromagnetic-ferromagnetic
transition at \(27\) kbar.\citep{song_switching_2019} It can also induce
large enhancements of magnetic interactions, e.g.~the ordering
temperature, \(T_c\), increases by \(250\) K in \ce{FeCl2} with the
application of \(420\) kbar
(\(Q = \frac{1}{T_c} \frac{\mathrm{d}T_c}{\mathrm{d}P} = +2.4\) \%
kbar\(^{-1}\)),\citep{xu_magnetism_2003} and in \ce{NiI2} \(T_c\)
increases by \(230\) K when compressed to \(190\) kbar (\(Q = +1.6\) \%
kbar\(^{-1}\)),\citep{pasternak_pressure-induced_1990} as well an
enhancement of the helimagnetic
state.\citep{kapeghian_effects_2023, occhialini_signatures_2023}
Pressure can also induce qualitative changes in the electronic structure
of vdW magnets,\citep{rozenberg_mott_2014} such as pressure-induced
metal-insulator
transitions.\citep{rozenberg_pressure-induced_2009, pasternak_pressure-induced_2001}
\ce{NiI2} undergoes a pressure-induced metal-insulator transition at
\(190\) kbar,\citep{pasternak_pressure-induced_1990} as does the
recently reported vdW antiferromagnet \ce{FePS3} at \(140\)
kbar.\citep{coak_emergent_2021} Pressure can even induce
superconductivity in \ce{FePSe3} above \(90\)
kbar.\citep{wang_emergent_2018}

Molecular frameworks, made from metals and molecular ligands, have
inherently higher flexibility than non-molecular materials comprising of
only atomic ions. The increased length of the ligands enhances the
extent of flexing of the
frameworks.\citep{serra-crespo_experimental_2015, wu_exceptional_2013}
They therefore can be expected to produce large responses at pressures
closer to those realisable in practical devices. For example, the
magnetic ordering temperature of the three-dimensional cyanide
frameworks, \ce{[Mn(4-dmap)]3[Mn(CN)6]2} (4-dmap =
4-dimethylaminopyridine) (\(Q = +13\) \%
kbar\(^{-1}\))\citep{kaneko_interpenetrated_2008} and
\ce{[Ru2(O2CCH3)4]3[Cr(CN)6]} (\(Q = +6.5\) \%
kbar\(^{-1}\))\citep{shum_observation_2007} rapidly increase with
pressure. Our understanding of the pressure response of the magnetism of
vdW molecular framework magnets is rapidly
developing.\citep{lopez-cabrelles_isoreticular_2018, Bassey2020, Perlepe2020}

\begin{figure}
  \hypertarget{fig:intro}{%
  \includegraphics[width=8.2cm]{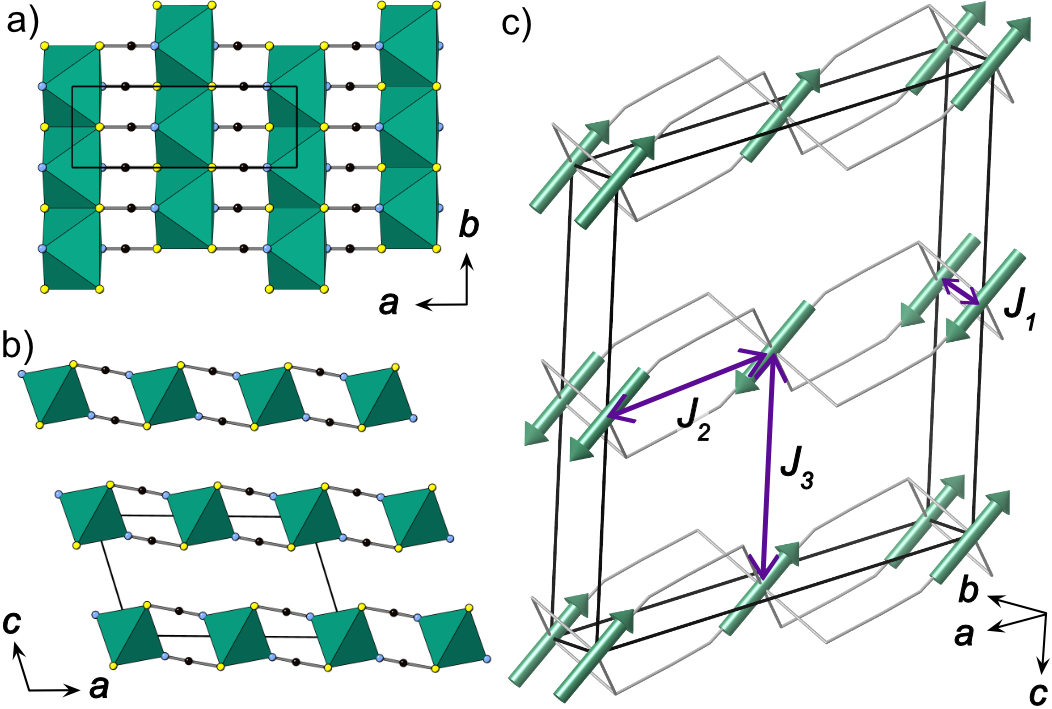}
  \caption{Structure of \ce{Ni(NCS)2} adapted from ref \protect\cite{Bassey2020} a) in plane structure, b) layer stacking. Ni=green octahedra, N=blue, C=black, S=yellow. c) The magnetic structure with the three nearest neighbour interactions identified. The \ce{Ni(NCS)2} framework is shown as a wireframe for clarity.}\label{fig:intro}
  } 
\end{figure}

Nickel(II) thiocyanate, \ce{Ni(NCS)2}, is a binary pseudohalide and the
only member of the \ce{M(NCS)2} family (M=Mn, Co, Fe, Ni, Cu) thus far
reported to have in-layer
ferromagnetism\citep{Bassey2020, Shurdha2012, dubler_intermediates_1982, Cliffe2018}
(Néel temperature, \(T_\mathrm{N} = 54\) K, \emph{c.f.} \ce{NiBr2}
\(T_\mathrm{N} = 52\) K\citep{adam_neutron_1980}), and is predicted to
be a single-layer ferromagnet.\citep{wu_nincs2_2021} It adopts an
analogous structure to that of the two-dimensional transition metal
halides \ce{MX2}, comprising \ce{NiS4N2} octahedra which edgeshare to
form layers in the \emph{ab} plane (Fig.~\ref{fig:intro}), which stack
along the \emph{c} direction. The rod shape and directionality of the
\ce{NCS^{-}} lowers the symmetry from the rhombohedral symmetry of
\ce{NiX2} to the monoclinic space-group \(C2/m\),\citep{Bassey2020}
which relieves potential frustration and means, unlike \ce{NiBr2} and
\ce{NiI2}, that no helimagnetic state is found.\citep{adam_neutron_1980}

Here we report the structural and magnetic changes of \ce{Ni(NCS)2} as
it is compressed to pressures up to \(8.4\) kbar. X-ray and neutron
powder diffraction experiments have allowed us to follow the evolution
of both lattice parameters and the structure under compression. Our
combination of magnetometry and low temperature neutron diffraction
measurements show the enhancement of the magnetic ordering temperature
with pressure and confirm the magnetic ground state throughout. We carry
out density functional theory (DFT) calculations which explain the
energetic origins of the observed behaviour.

\hypertarget{methods}{%
\section{Methods}\label{methods}}

\hypertarget{synthesis}{%
\subsection{Synthesis}\label{synthesis}}

The synthesis of the samples of \ce{Ni(NCS)2} was carried out following
the reported synthetic method of Bassey \emph{et al}.\citep{Bassey2020}
A typical synthesis (quantities as for the neutron sample) is described
below.

\ce{NiSO4.6H2O} (\(16.56\) g, \(63\) mmol) was dissolved in deionised
\ce{H2O} (\(50\) mL), giving a clear green solution. An aqueous solution
of \ce{Ba(SCN)2.3H2O} (\(19.37\) g, \(63\) mmol, \(120\) mL) was added,
with the rapid formation of a white precipitate and a green solution.
The reaction mixture was stirred at room temperature overnight and the
precipitate was removed by centrifugation and filtering under reduced
pressure. The solution was removed \emph{in vacuo}, giving a green-brown
microcrystalline powder of \ce{Ni(NCS)2}. The compound is stable to
humidity in the investigated conditions.

\hypertarget{magnetic-measurements}{%
\subsection{Magnetic Measurements}\label{magnetic-measurements}}

Measurements of the magnetic susceptibility were carried out on a
pelletised powder sample of \ce{Ni(NCS)2} using a Quantum Design
Magnetic Property Measurements System (MPMS) 3 Superconducting Quantum
Interference Device (SQUID) magnetometer with moment measurements
carried out in direct current (DC) mode. The measured data were
collected at pressures of \(1.2, \; 3.8, \; 5.2, \; 8.4\) kbar. The
zero-field-cooled (ZFC) susceptibility was measured in an applied field
of \(0.01\) T over the temperature range \(2\) to \(300\) K. The
pressure was applied using a BeCu piston cylinder pressure cell from
CamCool Research Limited, with a Daphne \(7373\) oil pressure medium. A
small piece of Pb was included in the sample space to act as a pressure
gauge.\citep{eiling_pressure_1981} As \emph{M}(H) is linear in this
field regime, the small-field approximation for the susceptibility,
\(\chi (T) \simeq \frac{M}{H}\), where \(M\) is the magnetisation and
\(H\) is the magnetic field intensity, was taken to be valid. Isothermal
magnetisation measurements were carried out on the same sample at \(10\)
K over the field range \(-7\) to \(+7\) T for each pressure point.

\hypertarget{dft}{%
\subsection{DFT}\label{dft}}

We have performed density functional theory calculations to probe the
structures and energetics of spin order of \ce{Ni(NCS)2}. Spin-polarized
DFT+\(U\) method (with Grimme's D3 van der Waals
correction\citep{grimme_consistent_2010}) was employed in structural
optimisations and energy calculations, using the Vienna Ab initio
Simulation Package (VASP)\citep{kresse_efficient_1996}. In our DFT+\(U\)
calculations, we used a \(U\) value of \(5.1\) eV for the d-electrons of
Ni\(^{2+}\)
cations,\citep{pickett_reformulation_1998, zhou_first-principles_2004}
and a range of ferromagnetic and anti-ferromagnetic spin solutions were
considered for magnetic Ni\(^{2+}\) cations (see Table 2, Supplemental
Material\citep{geers_high_2023}). We used a plane-wave basis set with a
kinetic energy cutoff of \(520\) eV to expand the wavefunctions. The
Perdew-Burke-Ernzerhof functional\citep{perdew_generalized_1996} in
combination with the projector augmented wave
method\citep{blochl_projector_1994, hobbs_fully_2000} were used to solve
the Kohn-Sham equations. An energy convergence threshold of \(10^{-6}\)
eV was used for electronic energy minimisation calculations, and the
structural optimisations, including cell parameters and atomic
positions, were considered converged if all interatomic forces fall
below \(0.01\) eV/Å. All DFT calculations have been performed in the
primitive cell (\(2\) formula units per cell) using a \emph{k}-grid with
a \emph{k}-points spacing of around \(0.1\) Å\(^{-1}\). To improve the
accuracy of the three nearest neighbour magnetic interactions determined
from DFT calculations at different pressures, we additionally performed
single point DFT energy calculations of the various spin configurations
at the optimised structures with a higher planewave cutoﬀ energy of
\(800\) eV. We note the relative energy difference between several spin
configurations in our DFT calculations can be less than \(1\) meV/metal
(depending on pressure), which is close to the DFT accuracy that we can
achieve with our current computational settings.

\hypertarget{synchrotron-diffraction-measurements}{%
\subsection{Synchrotron diffraction
measurements}\label{synchrotron-diffraction-measurements}}

X-ray powder diffraction experiments were performed at beamline I15 at
the Diamond Light Source, UK, with a wavelength of \(\lambda = 0.4246\)
Å, applying a custom-made high pressure powder X-ray diffraction setup
suitable for measurements up to \(4\) kbar.\citep{brooks_automated_2010}
A powder sample of \ce{Ni(NCS)2} was filled into a soft plastic
capillary together with silicone oil AP-100 as non-penetrating pressure
transmitting medium to maintain hydrostatic conditions. The capillaries
were sealed with Araldyte-2014-1. The capillary was loaded into, and in
direct contact with, the sample chamber consisting of a metal block
filled with water. The water acts as a pressure transmitting medium,
controlled with a hydraulic gauge pump.

Constant wavelength powder X-ray diffraction data were collected at room
temperature in the pressure range \(0.001\) to \(4\) kbar with a step of
\(0.2\) kbar and an estimated error of \(\pm 0.0030\) bar. The data were
processed using DAWN\citep{filik_processing_2017} with a \ce{LaB6}
calibration. Rietveld refinements of the nuclear model were completed
using the FullProf program\citep{rodriguez-carvajal_recent_1993} and
strain analysis carried out using PASCal.\citep{cliffe_pascal_2012}

\hypertarget{neutron-diffraction-measurements}{%
\subsection{Neutron Diffraction
Measurements}\label{neutron-diffraction-measurements}}

\hypertarget{high-resolution-measurements-d2b}{%
\subsubsection{High resolution measurements
(D2b)}\label{high-resolution-measurements-d2b}}

Constant wavelength neutron powder diffraction data were collected on
the high-resolution D2b diffractometer\citep{hewat_d2b_1986} at the
Institut Laue-Langevin (ILL), France. The incident wavelength was
\(\lambda = 1.59\) Å and the scattering was measured over an angular
range of \(10 < 2\theta < 160\)°. The sample was loaded in an aluminium
holder and placed within an aluminium gas pressure cell with helium used
as a pressure transmitting medium. Diffraction data were collected at
room temperature with applied pressures of \(1.7, \; 3.4, \; 5.1\) and
\(6.7\) kbar, with an estimated error of \(\pm 0.5\) kbar. Further
diffraction data were collected at a pressure of \(6.7\) kbar at
temperatures of \(20, \; 40\) and \(180\) K. The pressure was set at
room temperature, then the system was cooled for the low temperature
data. The cell was operated with a gas compressor to regulate the
pressure as the temperature was lowered and ensure maintenance of
constant pressure. A heater inside the gas capillary ensured the helium
was gaseous at low temperatures. NOMAD software\citep{mutti_nomad_2011}
from the ILL was used for data collection. Refinements of the nuclear
models were completed using the FullProf
program.\citep{rodriguez-carvajal_recent_1993}

\hypertarget{high-intensity-measurements-d1b}{%
\subsubsection{High intensity measurements
(D1b)}\label{high-intensity-measurements-d1b}}

Constant wavelength powder neutron diffraction data were collected on
the high-intensity medium resolution D1b
diffractometer\citep{orench_new_2014} at the Institut Laue Langevin
(ILL), France. The incident wavelength was \(\lambda = 2.52\) Å and the
scattering was measured over an angular range of \(2 < 2\theta < 128\)°.
The sample was loaded in an aluminium holder and placed within an
aluminium gas pressure cell with helium used as a pressure transmitting
medium. Diffraction data were collected at room temperature with applied
pressures of \(3, \; 4.5\) and \(6.7\) kbar, with an estimated error of
\(\pm 0.5\) kbar. A thermal diffractogram was collected at \(3\) kbar,
heated with a programmed ramp of \(0.05\) K min\(^{-1}\) between \(34\)
and \(65\) K. A second thermal diffractogram was collected at \(4.5\)
kbar, heated with a programmed ramp of \(0.07\) K min\(^{-1}\) between
\(44\) and \(60\) K. At a pressure of \(6.7\) kbar, data were collected
at a constant temperature of \(20\) K. The pressure was set at room
temperature, then the system was cooled for the low temperature data.
The cell was operated with a gas compressor to regulate the pressure as
the temperature was lowered and ensure maintenance of constant pressure.
A heater inside the gas capillary ensured the helium was gaseous at low
temperatures. NOMAD software\citep{mutti_nomad_2011} from the ILL was
used for data collection. Refinements of the nuclear and magnetic model
were completed using the FullProf
program.\citep{rodriguez-carvajal_recent_1993}

\hypertarget{results}{%
\section{Results}\label{results}}

\begin{figure}
  \hypertarget{fig:refinements}{%
  \includegraphics[width=8.2cm]{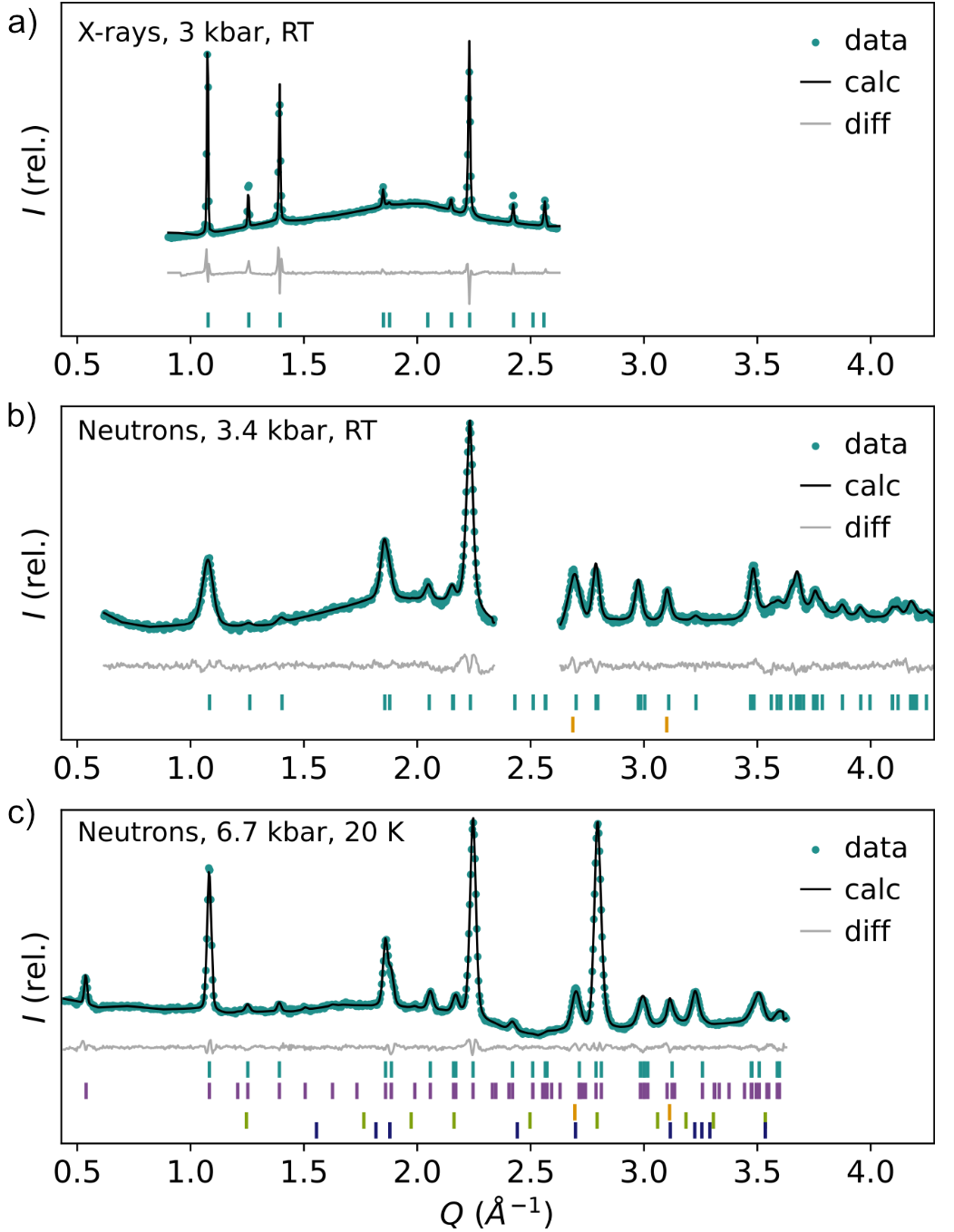}
  \caption{Rietveld refinements of diffraction data. a) Synchrotron X-ray data (I15, Diamond) measured at $3$ kbar and at ambient temperature, b) neutron data (D2b, ILL) measured at $3.4$ kbar and at ambient temperature, c) neutron data (D1b, ILL) measured at $6.7$ kbar and at $20$ K. Le Bail fits were made to account for reflections arising from the aluminium, oxygen and nitrogen. The tickmarks show the position of structural reflections (turquoise), magnetic reflections (purple), the aluminium sample holder (orange), oxygen (light green) and nitrogen (dark blue) phases.}\label{fig:refinements}
  } 
\end{figure}

The structural and magnetic changes that \ce{Ni(NCS)2} undergoes with
temperature at ambient pressure have previously been determined by
Bassey \emph{et al.}\citep{Bassey2020} The compound crystallises in the
monoclinic \(C2/m\) space group as two-dimensional layers stacked along
the \emph{c} axis. It has an ambient magnetic ordering temperature,
\(T_\mathrm{N}^0\), of \(54\) K, below which it magnetically orders as
an antiferromagnet. The moments correlate ferromagnetically within the
\emph{ab} plane and antiferromagnetically between adjacent planes,
ordering with a propagation vector of \textbf{k}\(=(0,0, \frac{1}{2})\)
in the magnetic space group \(C_c2/c\). The moments are oriented along
the \ce{N-Ni-N} bond direction and are restricted to the \emph{ac} plane
by symmetry. The three nearest neighbour interactions are: \(J_1\)
through \ce{Ni-S-Ni} (\emph{b} direction), \(J_2\) along \ce{Ni-NCS-Ni}
(\emph{a} direction) and \(J_3\) occurs between the metals on adjacent
layers (\emph{c} direction, Fig.~\ref{fig:intro}\(\mathrm{c}\)).

\hypertarget{nuclear-structure-under-pressure}{%
\subsection{Nuclear structure under
pressure}\label{nuclear-structure-under-pressure}}

\begin{table}
\caption{Compressibilities ($K$) for the principle axes of \ce{Ni(NCS)2} calculated from synchrotron X-ray data (I15, Diamond) at $4$ kbar and from neutron data (D2b, ILL) at $6.7$ kbar. The definition of the principal axes ($X_i$) are given relative to the unit cell ($a$, $b$, $c$) for the structure at $4$ kbar.}
\label{table:I15_strain}
\begin{tabular}{lrrrrr}
Axes  &\multicolumn{2}{c}{$K$ (TPa$^{-1}$)}    & $a$       & $b$    &  $c$ \\
         &     X-rays        &  Neutrons          &          &         &     \\
\hline \\
$X_1$    &        $3.8(4)$   &   $3(2)$          & $0.990$    & $0.0$  & $0.141$ \\
$X_2$    &        $13.5(1)$  &   $13.9(8)$       & $0.0$     & $1.0$  & $0.0$  \\
$X_3$    &        $32.5(2)$  &   $26.9(6)$       & $0.124$   & $0.0$  & $0.992$ \\
$V$      &        $50.0(5)$  &   $48(1)$         &           &        &    
\end{tabular}

\end{table}

\begin{figure}
  \hypertarget{fig:VP_RT_strain}{%
  \includegraphics{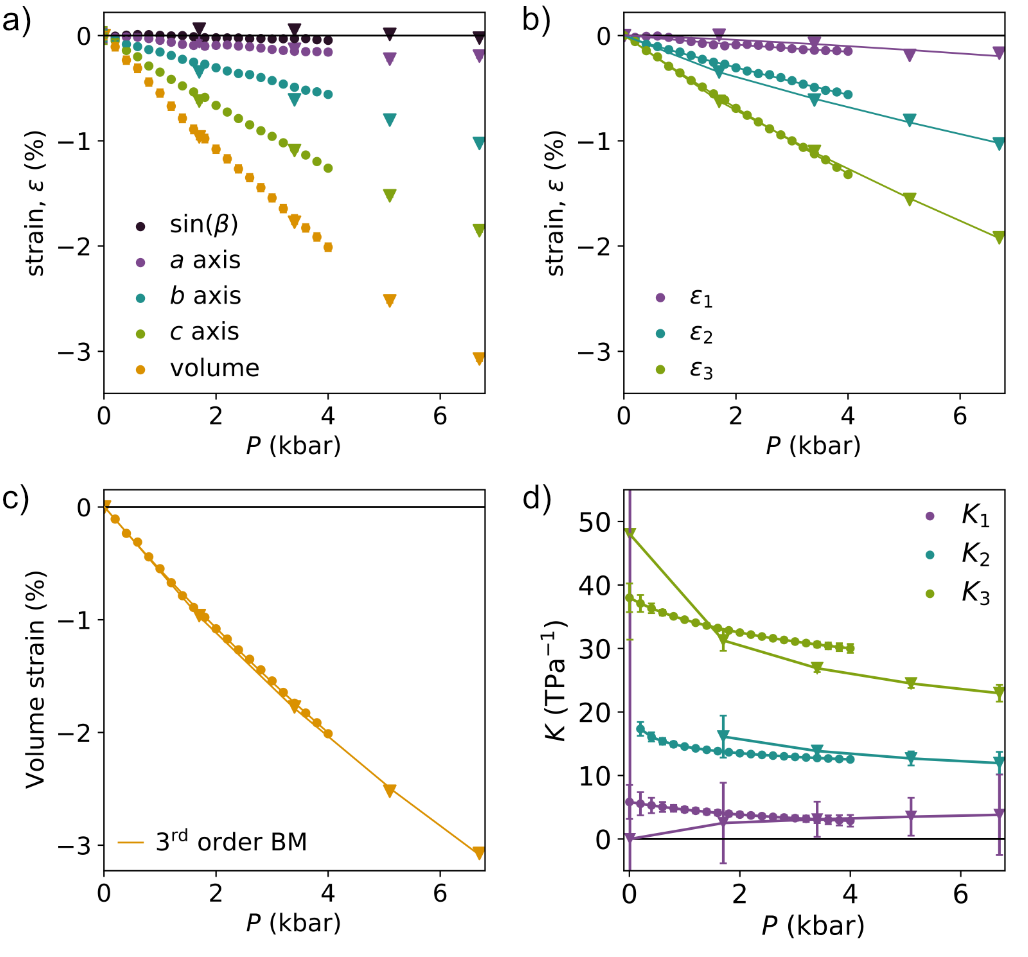}
  \caption{Crystallographic parameter variations determined from Rietveld refinements at ambient temperature using X-rays (circles, measured with I15, Diamond, $0.001$ to $4$ kbar) and neutrons (traingles, measured with D2b, ILL, $0.001$ to $6.7$ kbar). a) Normalised change of the unit cell parameters. b) Strain ($\varepsilon _i$) along the principal axes ($X_i$) with lines showing the linear compressibilities. c) The volume dependence with pressure, fitted with the third-order Birch-Murnaghan (BM) equation of state with $B_0=170(2)$ kbar and $B'=15(1)$ at $4$ kbar for X-rays and $B_0=156(6)$ kbar and $B'=22(3)$ at $6.7$ kbar for neutrons. d) The compressibility ($K$) along the principal axes as a function of pressure. Standard errors are shown with vertical lines.}\label{fig:VP_RT_strain}
  } 
\end{figure}

Ambient temperature diffraction measurements were carried out on a
powder sample of \ce{Ni(NCS)2} using X-rays (I15, Diamond, UK) between
\(0.001\) and \(4.0\) kbar (in \(0.2\) kbar steps) and with neutrons
(D2b, ILL, France) at \(1.7, \; 3.4, \; 5.1\) and \(6.7\) kbar. The
variation in lattice parameters was determined by performing Rietveld
refinements for both the X-ray and neutron data
(Fig.~\ref{fig:refinements}). The atom positions and anisotropic
displacement parameters were fixed to that of the structure at ambient
pressure.\citep{Bassey2020} Throughout the refinements the lattice
parameters were refined freely, along with background and peak shape
parameters. From these analyses we found no evidence of a phase
transition, with \ce{Ni(NCS)2} retaining its monoclinic symmetry
throughout the pressure regime. From refinements of the X-ray data, the
volume is reduced from \(225.18(6)\) Å\(^3\) at \(0.001\) kbar to
\(220.65(6)\) Å\(^3\) at \(4.0\) kbar. Fitting the X-ray data to a
third-order Birch-Murnaghan equation of state,\citep{birch_finite_1947}
gives a bulk modulus \(B_0=170(2)\) kbar and \(B'=15(1)\). Comparable
trends are observed for the compressibilities and reduction in axes
observed for the neutron data up to \(6.7\) kbar as well (\(B_0=156(6)\)
kbar and \(B'=22(3)\), Fig.~\ref{fig:VP_RT_strain}).

Both data sets, X-ray and neutron, show that as \ce{Ni(NCS)2} is
compressed, all unit cell axes decrease in length while the \(\beta\)
angle remains broadly constant (Fig.~\ref{fig:VP_RT_strain}). To assess
the degree of anisotropy, the strains along the principal axes were
calculated (Table \ref{table:I15_strain}) using the PASCal
program.\citep{cliffe_pascal_2012} The principal axes \(X_1\), \(X_2\)
and \(X_3\) lie in the following directions: \(X_1 \approx a\), along
the \ce{Ni-NCS-Ni} pathway; \(X_2=b\), along the \ce{Ni-S-Ni} bonds;
\(X_3 \approx c\), between the layers. The pressure dependence of the
principal axes shows the compressibilities,
\(K_i = -1/\varepsilon _i \frac{\mathrm{d}\varepsilon _i}{\mathrm{d}P}\),
derived from fitting an empirical equation of state
(\(\varepsilon _i (P)= \varepsilon _0 + \lambda(P-P_C) ^ \nu\)) are very
anisotropic. \(K_3 = 32.5(3)\) TPa\(^{-1}\), is more than double
\(K_2=13.5(1)\) TPa\(^{-1}\), and \(K_1=3.8(4)\) TPa\(^{-1}\) is an
order of magnitude smaller.

To explore the changes in the structure over a broader pressure range,
DFT calculations were performed at nominal pressures of \(0\), \(5\),
\(10\), \(20\), \(50\) and \(100\) kbar. These calculations, as they
were carried out at \(0\) K, were in the experimentally-derived layered
antiferromagnetic ground-state. The fitted bulk modulus using the
third-order Birch-Murnaghan equation of state is \(255.7(2)\) kbar, with
\(B' = 6.42(9)\). The compressibilities at \(10\) kbar are
\(K_3 = 18.1(5)\) TPa\(^{-1}\), \(K_2 = 10.1(4)\) TPa\(^{-1}\) and
\(K_1 = 3.0(2)\) TPa\(^{-1}\), broadly consistent with the measured
compressibilities. On increasing the pressure to \(100\) kbar we find,
as expected, a significant stiffening \(K_3 = 5.1(5)\) TPa\(^{-1}\),
\(K_2=5.5(8)\) TPa\(^{-1}\) \(K_1= 1.9(5)\) TPa\(^{-1}\).

\begin{figure}
  \hypertarget{fig:VT_CP_strain}{%
  \includegraphics{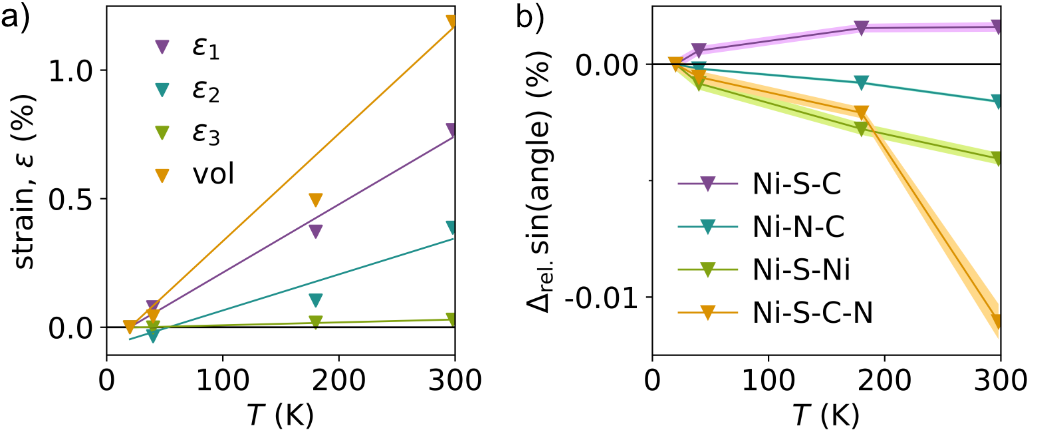}
  \caption{Selected data from Rietveld refinements at $6.7$ kbar measured at $20, \; 40, \; 180$ and $298$ K using the D2b diffractometer (ILL). a) Relative change in the principle axes and volume, b) relative change in sin(angle) of \ce{Ni-ligand} bonds and the dihedral angle \ce{Ni-S-C-N}. The shaded regions show the calculated errors.}\label{fig:VT_CP_strain}
  } 
\end{figure}

\hypertarget{variable-temperature-high-pressure}{%
\subsection{Variable temperature high
pressure}\label{variable-temperature-high-pressure}}

We also sought to investigate the effect of high pressure on the thermal
expansion of this material by measuring variable-temperature neutron
diffraction patterns (\(T=20\), \(40\), \(180\) and \(298\) K) at our
maximum pressure of \(6.7\) kbar. We find normal volumetric thermal
expansion, with a coefficient of thermal expansion
\(\alpha_V = 1/V \frac{\mathrm{d}V}{\mathrm{d}T}= 42(3)\) MK\(^{-1}\).
At \(6.7\) kbar, the thermal expansion is anisotropic with linear
thermal expansivities along the principal strain directions of
\(\alpha_1= 1/\varepsilon_1 \frac{\mathrm{d}\varepsilon_1}{\mathrm{d}T} = 26(1)\)
MK\(^{-1}\), \(\alpha_2= 14(2)\) MK\(^{-1}\) and \(\alpha_3 = 1.12(5)\)
MK\(^{-1}\) (broadly in the same directions as the compressibilities,
Fig.~\ref{fig:VT_CP_strain}\(\mathrm{a}\)). There are also significant
structural changes: with \(\angle\)\ce{Ni-S-Ni} (along \(X_2\))
expanding significantly on cooling,
\(\Delta \mathrm{sin(Ni-S-Ni)} = -0.0041(2)\) \%, whereas
\(\angle\)\ce{Ni-N-C} and \(\angle\)\ce{Ni-S-C} (predominantly along
\(X_1\)) change much less \(\Delta \mathrm{sin(Ni-N-C)} = -0.00160(6)\)
\% and \(\Delta \mathrm{sin(Ni-S-C)} = +0.0016(2)\) \%. The torsion
angle between \ce{Ni-S-C-N} decreases by
\(\Delta \mathrm{sin(Ni-N-C)} = -0.0111(7)\) \%
(Fig.~\ref{fig:VT_CP_strain}\(\mathrm{b}\)).

At this highest pressure, \(6.7\) kbar, additional peaks emerged in the
neutron datasets at high \(Q\) at temperatures below approximately
\(55\) K, which is well into the magnetic ordered phase. These likely
arise from a combination of oxygen and/or nitrogen phases crystallising
due to the presence of air in the pressure cell, and we accounted for
them using additional Le Bail
phases.\citep{tassini_high_2005, gorelli_crystal_2002} We found no
anomalies in the magnetometry data corresponding to this transition,
further supporting our assumption that this is not sample related.

\hypertarget{magnetometry}{%
\subsection{Magnetometry}\label{magnetometry}}

\begin{figure}
  \hypertarget{fig:susceptibility}{%
  \includegraphics[width=8.2cm]{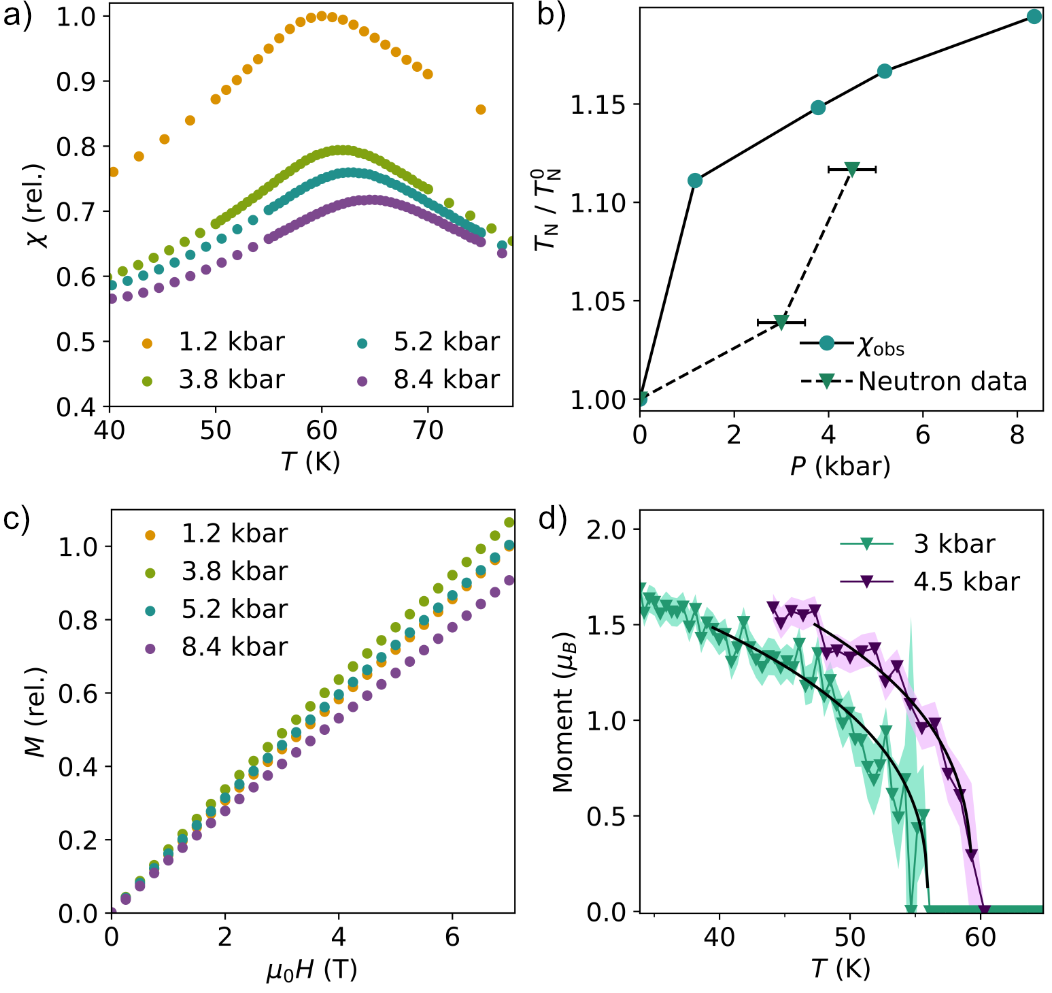}
  \caption{a) Magnetic susceptibility measured at $1.2(1)$, $3.8(1)$, $5.2(1)$ and $8.4(1)$ kbar, the relative susceptibility has been normalised to the maximum susceptibility at $1.2$ kbar. b) Relative change in $T_\mathrm{N}$ as a function of pressure derived from susceptibility (circles) and neutron diffraction data (triangles). Error bars related to pressure determination are shown for the neutron data, the error bars are contained within the symbols for the magnetometry data. The ordering temperatures were normalised to the ambient pressure ordering temperature, $T_\mathrm{N}^0=54$ K. c) Isothermal magnetisation measurements carried out at $10$ K plotted between $0$ to $+7$ T, showing the relative magnetisation normalised to the maximum magnetisation at $1.2$ kbar. d) The magnetic moment determined from Rietveld refinement of neutron diffraction data (D1b, ILL) at $3$ and $4.5$ kbar, fitted by a power law Eqn. \ref{eqn:tn_powerlaw} (black line). The shaded regions show the calculated errors of the magnitude of the moment.}
  \label{fig:susceptibility}
  } 
\end{figure}

To assess how the bulk magnetic properties of the material vary with
pressure, we carried out zero-field cooled susceptibility measurements
at \(1.2(1)\), \(3.8(1)\), \(5.2(1)\) and \(8.4(1)\) kbar on a
pelletised polycrystalline sample. At each pressure, the magnetic
susceptibility increases on cooling until a broad maximum is reached at
the ordering temperature (Fig.~\ref{fig:susceptibility}\(\mathrm{a}\)).
The ordering temperature shifts to higher values as the sample is
compressed, and is accompanied by a decrease in the maximum
susceptibility. At the highest measured pressure, \(8.4(1)\) kbar,
\ce{Ni(NCS)2} orders at \(T_\mathrm{N}= 64.6(4)\) K, a \(+19\) \%
increase from ambient pressure, giving \(Q = +2.3\) \% kbar\(^{-1}\)
(Fig.~\ref{fig:susceptibility}\(\mathrm{b}\)).

We also carried out isothermal measurements at the same pressure points
at \(10\) K between \(-7\) and \(+7\) T
(Fig.~\ref{fig:susceptibility}\(\mathrm{c}\)), which did not show
saturation in this field regime, as expected for a bulk antiferromagnet.
We found that the maximum magnetisation measured increased up to
\(3.8(1)\) kbar, before decreasing at higher pressures.

\hypertarget{magnetic-structure}{%
\subsection{Magnetic structure}\label{magnetic-structure}}

Our bulk magnetic measurements showed a significant enhancement in
magnetic ordering temperature, however to understand the detailed
evolution of the magnetic structure we carried out neutron diffraction
measurements on a powder sample (\(1.2\) g) using the high flux D1b
diffractometer at the ILL. Neutron diffraction data were collected below
the ordering temperatures at \(3, \; 4.5\) and \(6.7\) kbar at the
lowest temperatures for each pressure (\(P=3\) kbar, \(T=34\) K;
\(P=4.5\) kbar, \(T=44\) K; \(P=6.7\) kbar, \(T=20\) K). In these low
temperature data sets, we observed the appearance of an additional Bragg
reflection at \(0.54\) Å\(^{-1}\) at all pressures
(Fig.~\ref{fig:refinements}\(\mathrm{c}\)) and no other additional
reflections, which allowed us to confirm that the propagation vector is
\textbf{k}\(=(0, 0, \frac{1}{2})\), identical to that of the ambient
pressure material. This propagation vector corresponds to a doubling
along the \emph{c} axis, and we found that of the two maximal magnetic
space groups, only the ambient pressure structure, \(C_c2/c\), was able
to reproduce the observed intensity with a reasonable moment size.

We then carried out Rietveld refinements on these datasets in which the
magnitude of the nickel moment was fixed to \(1.75\; \mu_{\mathrm{B}}\),
in accordance with the reported size of the moment at \(2\) K at ambient
pressure,\citep{Bassey2020} due to the paucity of magnetic Bragg peaks.
This then allowed for the angle to be refined as the only free magnetic
parameter. At all three pressures, the angle of the moment remains
broadly unchanged from the ambient structure model, that is, the moment
is oriented along the \ce{N-Ni-N} bond within the \emph{ac} plane. As
the moment direction does not change up to \(6.7\) kbar or on warming at
ambient pressure,\citep{Bassey2020} for the following magnetic
refinements, therefore, we fixed the moment direction to point along the
\ce{N-Ni-N} bonds. This allowed us to refine the size of the magnetic
moment as the only free magnetic parameter in the last cycles of the
refinement.

Having established that the magnetic ground state remained unchanged, we
next looked to understand the nature of the ordering transition. We
collected variable temperature data at \(3\) kbar (\(T=34-65\) K) and
\(4.5\) kbar (\(T=44-60\) K). At \(3\) kbar, the magnetic moment was
refined to \(1.68(7) \; \mu_{\mathrm{B}}\) for the lowest temperature
point (\(T=34\) K) and the ordering temperature is found at
\(T_\mathrm{N} = 56.1(5)\) K. At \(4.5\) kbar the refined moment is
\(1.58(7) \; \mu_{\mathrm{B}}\) (\(T=44\) K) and decreases with
temperature to \(0 \; \mu_{\mathrm{B}}\) at \(T_\mathrm{N} = 58.0(5)\)
K. We were able to fit the refined magnetic moments to a power law in
the vicinity of the transition:

\begin{equation}\label{eqn:tn_powerlaw} 
M  = A (T_\mathrm{N} - T)^{\beta},
\end{equation}

\noindent where A is a proportionality constant, \(T_\mathrm{N}\) is the
ordering temperature and \(\beta\) is a critical exponent
(Fig.~\ref{fig:susceptibility}\(\mathrm{d}\)). For the fits, the errors
for A, \(T_\mathrm{N}\) and \(\beta\) are calculated from the square
root of the covariance matrix from the errors of the refined magnetic
moment. At \(3\) kbar, the fitted values were \(\beta=0.36(6)\) and
\(T_\mathrm{N} = 55.9(7)\) K. At \(4.5\) kbar, \(\beta=0.33(5)\) and
\(T_\mathrm{N} = 59.3(3)\) K. The values of \(\beta\) are in between
\(0.326\) and \(0.367\) expected for a three-dimensional Ising and
Heisenberg antiferromagnet.\citep{blundell_magnetism_2001}

\begin{figure}
  {%
  \includegraphics[width=8.2cm]{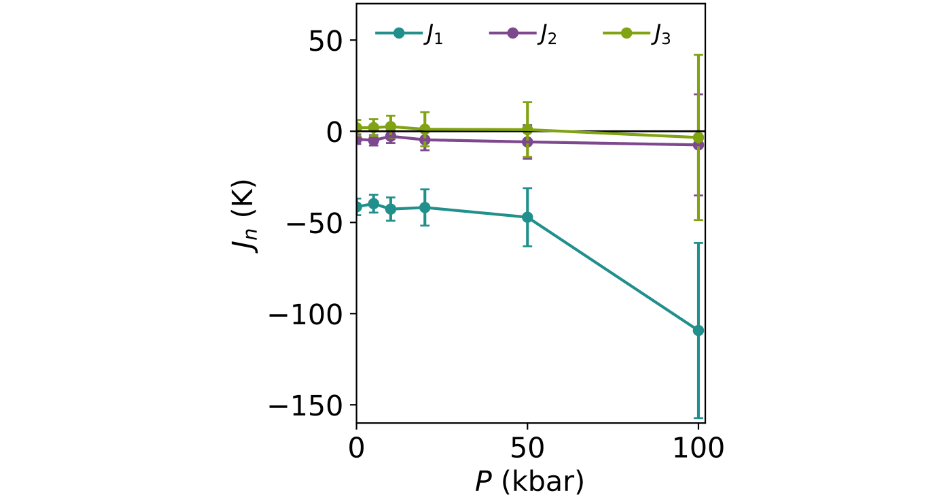}
  \caption{Calculated enthalpies of the three nearest neighbour interactions obtained from DFT calculations at ambient, $5, \; 10, \; 20, \; 50$ and $100$ kbar, where $J>0$ describes antiferromagnetic interactions and $J<0$ for ferromagnetic interactions. The calculated uncertainties associated with fitting to the Heisenberg Hamiltonian are shown with vertical lines.}\label{fig:DFT_Jvalue}
  } 
\end{figure}

To gain a deeper understanding of the exchange interactions responsible
for the magnetic behaviour we carried out DFT calculations on a variety
of spin states as a function of pressure. Geometry optimisations,
including both atomic positions and cell parameters, were carried out to
probe the magnetic ground state at the same pressures as previously
explored (\(0\), \(5\), \(10\), \(20\), \(50\) and \(100\) kbar), for
six high symmetry configurations (Table 3, Supplemental
Material\citep{geers_high_2023}), and the resultant energies and
enthalpies were calculated using the Hamiltonian
\(E = \sum_{ij} J_{ij} S_{i} \cdot S_j + E_0\).

We fitted these DFT-derived energies to a Heisenberg collinear spin
Hamiltonian to extract three interactions \(J_1\) along the \ce{Ni-S-Ni}
direction, \(J_2\) along the \ce{Ni-S-C-N-Ni} direction and \(J_3\),
corresponding the the interactions between layers. At ambient pressure,
\(J_1\) is large and ferromagnetic \(-43(8)\) K, \(J_2\) weaker and also
ferromagnetic \(-4(4)\) K and \(J_3\) is zero within error, consistent
with the expected ground state (Fig.~\ref{fig:DFT_Jvalue}). The large
values of the error suggest that the Heisenberg Hamiltonian we employed
is perhaps not appropriate: either due to large single ion effects or
higher order interactions (e.g.~biquadratic interactions). It is likely
that up to \(50\) kbar the reported ambient antiferromagnetic structure
is the most stable configuration. However, we did not find a strong
trend in the predicted \(J\) values with pressure beyond the error of
the calculations, since the relative energy differences between selected
spin configurations here are less than \(1\) meV/metal, meaning we are
close to the limits of accuracy for these DFT calculations.

\hypertarget{discussion}{%
\section{Discussion}\label{discussion}}

From our diffraction data, we have calculated the bulk modulus of
\ce{Ni(NCS)2} to be \(B_0=170(2)\) kbar, which shows that \ce{Ni(NCS)2}
is one of the softer van der Waals compounds, e.g.~for \ce{CrBr3}
\(B_0=230\) kbar\citep{ahmad_pressure-driven_2020} and \ce{FePSe3}
\(B_0=828\) kbar.\citep{wang_emergent_2018} High pressure structural
phase transitions are common in layered
compounds,\citep{haines_pressure-induced_2018, wang_pressure-induced_2017, coak_emergent_2021}
however our DFT calculations do not provide any evidence of a structural
phase transition up to \(100\) kbar. Ab initio structure searches and
larger supercell calculations together with higher pressure calculations
would allow further exploration of the potential for new \ce{Ni(NCS)2}
phases. These calculations also suggest that any metallisation
transition occurs significantly above \(100\) kbar, compared to
\ce{NiI2} \(P_c = 190\) kbar,\citep{pasternak_pressureinduced_1991}
\ce{FeCl2} \(P_c=450\) kbar\citep{rozenberg_pressure-induced_2009} and
\ce{FePS3} \(P_c \approx 140\) kbar.\citep{coak_emergent_2021}

\begin{figure}
  \hypertarget{fig:hinge}{%
  \includegraphics[width=8.2cm]{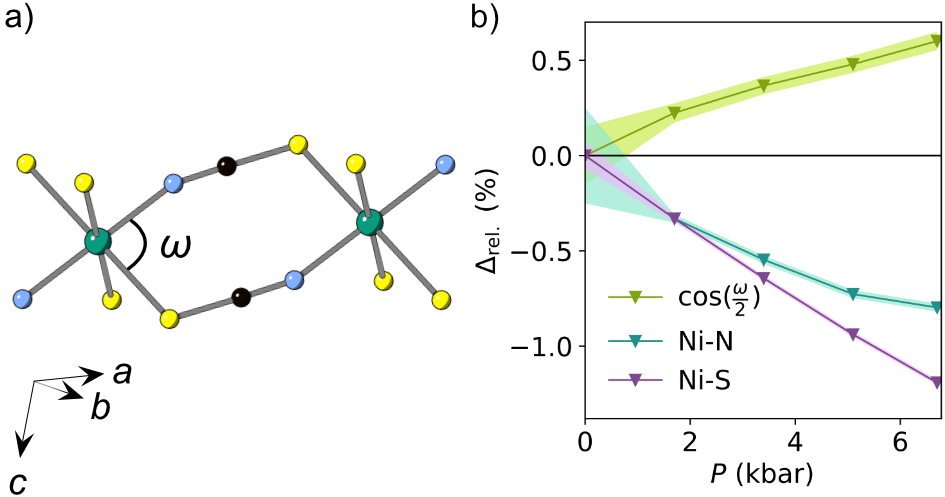}
  \caption{a) Nickel ions connected through two NCS ligands, showing the 'hinge' angle, $\omega$. Ni=green, N=blue, C=black, S=yellow. b) Relative change in the bond lengths \ce{Ni-N} and \ce{Ni-S}, and the relative change in $cos(\frac{\omega}{2})$ with pressure. The ambient pressure values were obtained from ref \protect\cite{dubler_intermediates_1982}. The shaded regions show the errors.}\label{fig:hinge}
  } 
\end{figure}

We can also compare \ce{Ni(NCS)2} with other pseudobinary molecular
frameworks. Since changing the metal cation species can influence the
flexibility of the structure, we focus here on \ce{Ni^{2+}}
frameworks.\citep{grover_tuning_2023} The three-dimensional
\ce{Ni(dca)2} (dca=\ce{N(CN)2}) has a larger bulk modulus, \(B_0=360\)
kbar,\citep{fan_negative_2019} than \ce{Ni(NCS)2}. As the ligand length
increases, it would be expected that the structure becomes more
flexible,\citep{dissegna_tuning_2018} however, as \ce{Ni(dca)2} is
three-dimensionally connected, the resultant bulk modulus is larger than
that of layered \ce{Ni(NCS)2}. The volumentric compressibility provides
only a partial picture, as both compounds are anisotropic. The
compressibility of \ce{Ni(dca)2} along the \emph{b} and \emph{c} axes is
similar to that of \ce{Ni(NCS)2} along the \(X_2\) and \(X_3\)
(\(b = 11.1\) TPa\(^{-1}\) and \(c = 24.1\) TPa\(^{-1}\)), but
\ce{Ni(dca)2} shows negative linear compressibility (NLC) in the third
(\emph{a}) direction. The materials have similar ligand arrangements
along the \emph{a} axis, where the nickel ions are bridged by two
ligands coordinated in an end-to-end arrangement. In both materials, the
`hinge' ligand\(-\)metal\(-\)ligand bond angles, \(\omega\)
(\(\omega = \angle\)\ce{N-Ni-S} and \(\angle\)\ce{N-Ni-N},
Fig.~\ref{fig:hinge}) decrease: \([\Delta\)
cos\((\frac{\omega}{2})]/P = 0.09\) \% kbar\(^{-1}\) (NCS), compared to
\([\Delta\) cos\((\frac{\omega}{2})]/P = 0.16\) \% kbar\(^{-1}\) (dca).
In \ce{Ni(NCS)2}, the softer \ce{Ni-L} bonds mean that there is still
contraction in the bond lengths (\ce{Ni-N}\(= -0.12\) \% kbar\(^{-1}\)
and \ce{Ni-S}\(=-0.18\) \% kbar\(^{-1}\)). This competition between the
two components is likely the cause of the apparent stiffness of the
\emph{a} axis. In contrast, the square-planar \ce{Ni(CN)2} is
significantly stiffer, \(B_0 = 1050\) kbar, perhaps due to the planarity
of the structure and linearity of \ce{Ni-CN-Ni}
bonds.\citep{mishra_new_2016}

Our high pressure variable temperature data show that, despite being a
layered structure, \ce{Ni(NCS)2} has near zero thermal expansion along
the \(X_3\) direction. In comparison, there is positive thermal
expansion along \(X_1\) and \(X_2\) directions
(Fig.~\ref{fig:VT_CP_strain}). Calculating the thermal expansion
coefficients from the published ambient pressure data\citep{Bassey2020}
gives values of \(\alpha _1 = 26(2)\) MK\(^{-1}\),
\(\alpha _2 = 2.1(9)\) MK\(^{-1}\) and \(\alpha _3 = -5.9(9)\)
MK\(^{-1}\). Here, the principal axes are defined as
\(X_1 = 0.47 a + 0.88 c\), \(X_2=b\) and \(X_3 = -0.56 a + 0.83 c\) at
\(100\) K. This is rotated slightly in comparison to the high pressure
data (definitions are approximately equal to directions given in Table
\ref{table:I15_strain}). \(X_3\) exhibits negative thermal expansion
which is uncommon, however it can be observed in other anisotropic
molecular framework
compounds.\citep{hodgson_negative_2014, collings_compositional_2016}
These values are comparable to those seen at \(6.7\) kbar, but the
within-layer expansion has decreased. This behaviour is not uncommon in
layered molecular frameworks and likely reflects the stiffening of the
transverse out-of-plane vibrations.\citep{hodgson_negative_2014} Since
the ambient pressure structure was measured between \(1\) and \(100\) K,
and our high pressure data were measured between \(20\) and \(298\) K,
direct comparison of coefficients should be done cautiously. Future
investigations would be needed for detailed quantitative comparisons.

The most significant change observed on compressing \ce{Ni(NCS)2} is the
marked increase in magnetic ordering temperature, \(Q = +2.3\) \%
kbar\(^{-1}\). This increase is large compared to other nickel molecular
frameworks, such as \ce{Ni(dca)2} (\(Q = +0.4\) \%
kbar\(^{-1}\)),\citep{nuttall_pressure_2000} \ce{NH4Ni(HCOO)3}
(\(Q = +0.2\) \% kbar\(^{-1}\))\citep{collings_pressure_2018} and
\ce{NiCl2(pym)2}, pym = pyrimidine, (\(Q = +1.3\) \%
kbar\(^{-1}\),\citep{kreitlow_pressure_2005} for additional examples see
\citep{pasternak_pressure-induced_1990, lis_structural_2023, vettier_magnetic_1975, rozenberg_pressure-induced_2009, xu_magnetism_2003, zhang_pressure-enhanced_2021, xu_unique_2020, coak_emergent_2021, wang_emergent_2018, nuttall_pressure_2000, fan_negative_2019, yakovenko_pressure-induced_2015, collings_pressure_2018, collings_structural_2016, clune_developing_2020, Collings2018, kreitlow_pressure_2005, wolter_pressure_2003}
summarised in the Supplemental Material\citep{geers_high_2023} Fig. 4
and Table 5) The increase of ordering temperatures in these compounds
results from both changes in exchange interactions, including the
increase in strength of interactions and reduction of frustrating
interactions, and the single-ion anisotropy. Our DFT calculations
suggest this compound is not magnetically frustrated, and so the general
anticipated enhancement of exchange will be largely reinforcing. The
significant uncertainties in the DFT-derived superexchange parameters
suggest that higher level calculations, incorporating single-ion
anisotropy and higher-order exchange interactions, might be necessary
for complete understanding of the underlying magnetic Hamiltonian. The
interlayer \(J_3\) pathway, through
\(\mathrm{Ni-S} \cdots \mathrm{S-Ni}\), decreases in distance
significantly, and is likely the limiting factor in the ordering
temperature.

This shift towards a three-dimensional exchange network can be seen in
the critical exponent of the temperature dependence of the staggered
magnetic moment. A power law fit (Eqn. \ref{eqn:tn_powerlaw}) of the
Rietveld-refined magnetic moment as a function of temperature at \(3\)
and \(4.5\) kbar (Fig.~\ref{fig:susceptibility}\(\mathrm{d}\)) gave a
critical exponents of \(\beta = 0.36(6)\) and \(0.33(5)\), comparable to
the three-dimensional Heisenberg results (\(\beta=0.367\)), whereas
fitting to the published ambient pressure data\citep{Bassey2020} gave
\(\beta = 0.25(4)\) (and \(T_\mathrm{N}=60.3(6)\) K). The high pressure
value is consistent with those of three-dimensional molecular framework
magnets, e.g.~\ce{Mn(dca)2(pyz)} (pyz=pyrazine) where
\(\beta = 0.38\),\citep{manson_long-range_2001} whereas the ambient
pressure exponent is closer to that of other vdW magnets, e.g \ce{NiCl2}
\(\beta = 0.27\)\citep{lindgard_spin-wave_1975} and \ce{FeCl2}
\(\beta=0.29\).\citep{yelon_magnetic_1972} Unfortunately, comparatively
large errors (relative to the small energy differences between selected
spin configurations) in our DFT-derived superexchange parameters do not
allow us to further buttress this picture (Fig.~\ref{fig:DFT_Jvalue}).
More direct exploration of the changes in the exchange parameters,
whether through spectroscopy or other bulk measurements, would be
valuable to explore this two-dimensional to three-dimensional crossover.

We find no evidence of any spin rotation with pressure in \ce{Ni(NCS)2}
up to \(6.7\) kbar. This is consistent with the behaviour of other vdW
metal halides, including \ce{NiI2},
\ce{CoI2}\citep{pasternak_pressureinduced_1991} and
\ce{FeCl2}.\citep{vettier_magnetic_1975} This is perhaps due to strong
easy-axis single-ion anisotropy in this
material.\citep{defotis_antiferromagnetism_1993, wu_nincs2_2021}

\hypertarget{conclusion}{%
\section{Conclusion}\label{conclusion}}

We have investigated the effects of pressure on the structure and
magnetism of the van der Waals framework \ce{Ni(NCS)2}. X-ray and
neutron powder diffraction data reveal strongly anisotropic strain with
the interlayer direction being an order of magnitude more compressible
than the \emph{a} direction. Low temperature neutron diffraction
measurements combined with susceptibility data show there is a
significant increase in the magnetic ordering temperature with pressure
driven by the reduction of interlayer separation. This work has explored
the use of a thiocyanate framework as a source of flexibility to enhance
the changes in the magnetic properties of a van der Waals material. It
suggests that further molecular framework magnets hold potential to show
marked responses to compression. Following this work, it would also be
worthwhile to explore the monolayer limits of \ce{Ni(NCS)2} under
pressure to observe if more drastic changes to the magnetic structure
can be obtained for a few-layer molecular framework.

\section*{Acknowledgements}
M.G. acknowledges the Institut Laue Langevin (ILL, France) Graduate
School for provision of a studentship. M.J.C. acknowledges the School of
Chemistry, University of Nottingham for support from Hobday bequest. We
acknowledge the ILL for beamtime under proposal number 5-24-660. The raw
data sets are available at https://doi.ill.fr/10.5291/ILL-DATA.5-24-660.
S.L. acknowledges the use of the Sulis supercomputer through the HPC
Midlands+ Consortium, and the ARCHER2 supercomputer through membership
of the UK's HPC Materials Chemistry Consortium, which are funded by
EPSRC grant numbers EP/T022108/1 and EP/R029431/1, respectively. The
MPMS measurements were carried out using the Advanced Materials
Characterisation Suite in the Maxwell Centre, University of Cambridge,
funded by EPSRC Strategic Equipment Grant EP1M000524/1. We acknowledge
Diamond Light Source (UK) for beamtime on beamline I15 under proposal
number CY30815-2.

\section*{Author contributions statement}
M.G. and M.J.C. synthesised the samples. M.J.C., D.M.J., C.L. and S.S.S.
carried out the magnetometry measurements and analysis. A.B.C., M.G.,
O.F., L.C.D. and M.J.C. carried out the neutron diffraction experiment
and analysis. S.L. carried out the density functional theory
calculations. J.P., E.M., S.A.H., S.M.K., G.K., and D.D. carried out the
X-ray diffraction measurements. M.J.C., M.G., L.C.D. performed the
analysis of the X-ray diffraction data. M.G., L.C.D. and M.J.C. wrote
the paper with contributions from all authors.

\bibliographystyle{apsrev4-1}
\bibliography{paper.bib}

\end{document}